\documentclass[final,authoryear,3p,times,twocolumn]{elsarticle}  % Pseudo published article view

\usepackage{graphicx}
\usepackage{amssymb}
\usepackage{amsmath}
\usepackage{float}
\usepackage{color}
\usepackage{natbib}
\usepackage{url}
\usepackage{subfig}
\usepackage{booktabs}
\usepackage{multirow}
\usepackage{gensymb}  % contains \degree and \celsius

\journal{Planetary \& Space Science}
\begin{document}
%%%%%%%%%%%%%%%%%%%%%%%%%%%%%%%%%%%%%%%%%%%%%%%%%
\begin{frontmatter}



\title{The Need for Speed in Near-Earth Asteroid Characterization}


%%%%%%%%%%%%%%%%%%%%%%%%
\author[a]{J.L.~Galache\corref{cor1}}
\ead{jlgalache@cfa.harvard.edu}
\author[b,c]{C.L.~Beeson}
\ead{charlie.beeson@sky.com}
\author[d]{K.K.~McLeod}
\ead{kmcleod@wellesley.edu}
\author[b]{M.~Elvis}
\ead{melvis@cfa.harvard.edu}
\cortext[cor1]{Corresponding author. Tel.: +1 617-495-7440.}

\address[a]{Minor Planet Center, Harvard-Smithsonian Center for Astrophysics, 60 Garden Street, Cambridge, 02138 MA, USA}
\address[b]{Harvard-Smithsonian Center for Astrophysics, 60 Garden Street, Cambridge, 02138 MA, USA}
\address[c]{School of Physics \& Astronomy, University of Southampton, Southampton, Hampshire, SO17 1BJ, UK}
\address[d]{Whitin Observatory, Wellesley College, Wellesley, 02481 MA, USA}

%%%%%%%%%%%%%%%%%%%%%%
\begin{abstract}
We have used Minor Planet Center (MPC) data and tools to explore the discovery circumstances and properties of the currently known population of over 10,000~NEAs, and to quantify the challenges for  follow-up from ground-based optical telescopes.  The increasing rate of discovery has grown to $\sim$1,000/year as surveys have become more sensitive, by 1~mag every $\sim$7.5~years. However, discoveries of large ($H\leq22$) NEAs have remained stable at $\sim$365/year over the past decade, at which rate the 2005 Congressional mandate to find 90\% of 140~m NEAs will not be met before 2030 (at least a decade late). Meanwhile, characterization is falling farther behind: Fewer than 10\% of NEAs are well characterized in terms of size, rotation periods, and spectral composition, and at the current rates of follow-up it will take about a century to determine them even for the known population. Over 60\% of NEAs have an orbital uncertainty parameter, $U\geq4$, making reacquisition more than a year following discovery difficult; for $H>22$ this fraction is over 90\%. We argue that rapid follow-up will be essential to characterize newly-discovered NEAs. Most new NEAs are found within 0.5~mag of their peak brightness and fade quickly, typically by 0.5/3.5/5~mag after 1/4/6~weeks. About 80\% have synodic periods of $<$3~years that would bring them close to Earth several times per decade. However follow-up observations on subsequent apparitions will be difficult or impossible for the bulk of new discoveries, as these will be smaller ($H>22$) NEAs that tend to return 100$\times$ fainter. We show that for characterization to keep pace with discovery would require: Quick (within days) visible spectroscopy with a dedicated $\geq$2~m telescope; long-arc (months) astrometry, to be used also for phase curves, with a $\geq$4~m telescope; and fast-cadence ($<$min) light curves obtained rapidly (within days) with a $\geq$4~m telescope. For the already-known large ($H\leq22$) NEAs, that tend to return to similar brightness, subsequent-apparition spectroscopy, astrometry, and photometry could be done with 1--2~m telescopes.
\end{abstract}
%%%%%%%%%%%%%%%%%%%%%%%


\end{frontmatter}

%%%%%%%%%%%%%%%%%%%%%%%% INTRODUCTION

\section{Introduction}
\label{introduction}

Near-Earth Asteroids (NEAs) are asteroids that have been brought into the inner Solar System mainly through gravitational interactions with Jupiter and Saturn, placing them in orbits with perihelia $q\leq1.3$~AU that intersect or come close to Earth's orbit \citep{Shoemaker1979,Bottke2002,Greenstreet2012a}. This population is of interest to science as it provides a probe into the dynamical and compositional evolution of our Solar System and, not least, as a source of objects that have shaped Earth's history, geologically and biologically, through numerous impacts over many millions of years. And they may yet do so again \citep{Shoemaker1983}. 

NEAs have attracted public interest in recent times due to the unexpected 2013 February 15 event when a $\sim$17~m meteor exploded with 500~kt TNT of energy over the city of Chelyabinsk, Russia and injured over 1,000 people \citep{Emelyanenko2013,Borovicka2013}. The start-up of two private companies aiming to mine valuable resources from NEAs in the near future has rekindled interest in the scientific and for-profit exploration of asteroids \citep[e.g.,][]{Elvis2014a}. In addition, the $44^{th}$ U.S. president, Barack Obama, has made NEAs the prime targets for human space exploration\footnote{\url{http://www.nasa.gov/news/media/trans/obama_ksc_trans.html}}. 

In 2005 the U.S. Congress issued a mandate to NASA to find at least 90\% of the NEAs larger than 140~m (corresponding to an absolute magnitude\footnote{for asteroids, the absolute magnitude $H$ is given by the apparent $V$ magnitude that the asteroid would have if it could be observed from 1~AU away, at zero phase angle, while it was 1~AU from the Sun.} of $H\leq22$) by 2020 \citep{Brown2005}. There are an estimated $13,200\pm1,900$ in this size range, and $20,500\pm2,000$ larger than 100~m, or $H\lesssim23$ \citep{Mainzer2011b}. Some argue that follow-up efforts aimed at characterization should concentrate only on the subset of NEAs that come close to Earth (Minimum Orbit Intersection Distance\footnote{The Earth MOID is the shortest distance separating the orbit of an asteroid from that of Earth. A small MOID value does not necessarily imply a risk of impact as both Earth and the asteroid are rarely at the points of their orbits closest to each other at the same time.}, $\rm MOID\leq0.05$~AU)  \textit{and} are larger than $\sim$140~m; these are formally classified as Potentially Hazardous Asteroids (PHAs)\footnote{\url{http://www.minorplanetcenter.net/iau/lists/Dangerous.html}}. However, there are good reasons to design follow-up programs that are broader than that. In terms of hazard assessment, we know that the much more numerous objects smaller than 140~m can pose significant risks \citep{Brown2013}, as we were reminded by the Chelyabinsk event.  A large majority ($\sim$67\%) of the $>$4,900 known NEAs with $H>22$ have $\rm MOID\leq0.05$~AU. In addition, because PHAs (irrespective of size) are, as far as we know, drawn from the same general population as other NEAs, characterizing the general population will allow us to calibrate the relationships needed to infer physical and orbital evolution properties of the PHA subset. Finally, from the technological standpoint, NASA's proposed Asteroid Redirect Mission\footnote{\url{http://www.kiss.caltech.edu/study/asteroid/asteroid_final_report.pdf}} requires an 8~m NEA in a favorable orbit be found. A second option for ARM is the retrieval of a 1--5~m boulder from a larger NEA \citep{Abell2014}.

As we argue below, most kinds of follow-up that would help to quantify an object's threat or usefulness must be done within a short window of time, before some of the properties are well-enough known to judge whether it is an ``important'' object for follow-up or not. Thus we concentrate our discussion on the NEA population as a whole.

In this paper, we use the discovery circumstances and properties of the currently known population of NEAs to quantify some of the challenges that ground-based observers face in making follow-up observations. Our goal is to define the optimal follow-up strategies for NEAs that would allow the bulk of the discovered population to be characterized on a one-decade timescale. This is desireable from a planetary defence perspective because the bulk of the impact hazard resides with the smaller ($H\leq22$), and thus more numerous, objects, which are also the least characterized. Gathering these data (composition, structure, size, etc.) for a significant proportion of this size population would allow for better estimates of potential damage due to impact, and more optimistically, better designed missions to deflect or destroy them. For example, if we knew that most 25~m NEAs are monolithic, we would produce different impact damage estimates and deflection mission requirements than if we discovered most of them are rubble piles. Scientifically, knowing the spectral class and spin distributions of a significant portion of a given size population will enable tests of current dynamical evolution models of the inner Solar System as well as possible insights into the collisional history of asteroids and the effect of Solar radiation pressure on spin axis evolution.

We describe our data sources in \S\ref{data_sources}, and present current discovery and characterization trends in \S\ref{current_nea}.  We then describe the known NEA population in terms brightness behavior (\S\ref{brightness}), positional uncertainties (\S\ref{positional_constraints}), sky motions (\S\ref{sky_motion}), and hemisphere bias (\S\ref{hemisphere_bias}). In \S\ref{discussion}, we discuss the implications for follow-up astrometric, photometric, and spectroscopic observations. Finally, in \S\ref{conclusions} we summarize by laying out strategies for increasing the rate of characterization.

In this paper, we use the discovery circumstances and properties of the currently known population of NEAs to quantify some of the challenges that ground-based observers face in performing follow-up observations. Our goals are to define the optimal follow-up strategies for NEAs so the bulk of the discovered population can be characterized on a one-decade timescale, and to provide a centralized source of NEA discovery properties useful for observers planning follow-up programs.

%%%%%%%%%%%%%%%%%%%%%%%%%%%%%%%%%%%% METHOD
\section{Method}
\label{data_sources}

The IAU Minor Planet Center\footnote{\url{http://www.minorplanetcenter.net}} (MPC) serves as the world's clearinghouse for NEA observations and orbital data, and provides web-based tools for generating lists of objects that are currently observable, and providing NEA ephemerides. We have written a series of Python programs making use of modules \texttt{Mechanize}\footnote{\url{http://wwwsearch.sourceforge.net/mechanize/download.html}} and \texttt{BeautifulSoup4}\footnote{\url{http://pypi.python.org/pypi/beautifulsoup4/}}, to extract data from the MPC webpages and files.  We used the following MPC resources.

\noindent Data:

\begin{itemize}
\item{\texttt{NEA.txt}\footnote{\url{http://www.minorplanetcenter.net/iau/MPCORB/NEA.txt}} provides orbital elements at a nominal epoch, along with absolute magnitude $H$, slope parameter $G$\footnote{The Slope Parameter, $G$, is a measure of how an object's brightness surges when it nears opposition, the so-called \textit{opposition effect}. It is believed to be an interplay of shadowing and coherent-backscattering mechanisms \citep[e.g.,][]{Muinonen2002}.}, number of observations, number of oppositions, arc length (for single opposition objects) or years of first and last observations (for multiple opposition objects), and orbital uncertainty parameter $U$ (see \S\ref{orbit_uncertainty} for an explanation of this parameter).}
\item{The lists of Atens\footnote{\url{http://www.minorplanetcenter.net/iau/lists/Atens.html}}, Apollos\footnote{\url{http://www.minorplanetcenter.net/iau/lists/Apollos.html}}, and  Amors\footnote{\url{http://www.minorplanetcenter.net/iau/lists/Amors.html}} provide much of the same data but also include discovery date and site, and Earth MOID, but exclude $U$.}
\end{itemize}

\noindent Tools:
\begin{itemize}
\item{\texttt{MPEph}\footnote{\url{http://www.minorplanetcenter.net/iau/MPEph/MPEph.html}} is used to generate a table of ephemerides for user-selected NEAs for a range of dates and times.  The data returned include $V$ magnitude estimates.}
\item{\texttt{NEAobs}\footnote{\url{http://scully.cfa.harvard.edu/cgi-bin/neaobs.cgi}} is used to generate a list of NEAs suitable for observation according to user-specified criteria including observing location, NEA magnitude, sky motion, solar elongation, RA, Dec, and orbital uncertainty.}
\end{itemize}

We have also made use of the web services of NEODyS-2\footnote{\url{http://newton.dm.unipi.it/neodys/}}, which provides convenient text tables of observations for each NEA taken from the MPC's database.

To quantify the various observational follow-up challenges, we use these resources to look at the properties of all NEAs discovered as of 2013 March as described in the sections below. For some of the analysis we also make use of a subsample consisting of the 6,763 NEAs discovered during the 10~years spanning January 1, 2002 to December 31, 2011.  We call this the ``Decade Sample.''  For each NEA in the Decade Sample, we generated a file containing a list of daily ephemerides for that time span, including calculated apparent $V$ magnitudes.

\section{Current NEA Characterization Progress Rates and Trends}
\label{current_nea}

\begin{figure}[!ht]
\begin{center}
\includegraphics[width=0.99\linewidth]{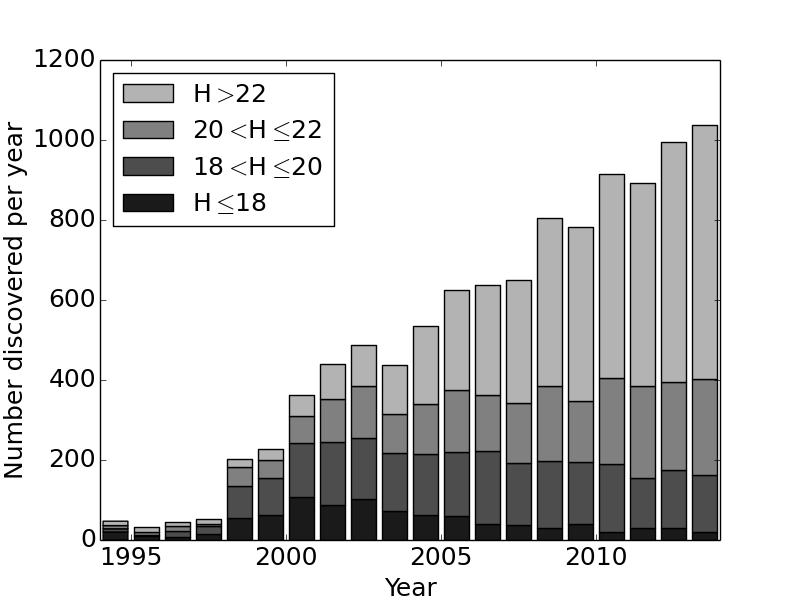}
\caption[Discovery rate of NEAs for 1994--2013.]{Discovery rate of NEAs for 1994--2013. Blue denotes $H\leq22$ and yellow, $H>22$.}
\label{disc_rate}
\end{center}
\end{figure}

The current discovery rate of NEAs is $\sim$1,000/year (Figure~\ref{disc_rate}) and this rate should increase shortly to $\gtrsim$1,500/year with the enhanced Catalina Sky Survey detector arrays being installed \citetext{E. Christensen, priv.\ comm.}, the increased use of Pan-STARRS-1 for NEA discovery (100\% of available time beginning March 2014) and the coming online of Pan-STARRS-2 \citetext{R. Wainscoat, priv.\ comm.}, and the first light of the LINEAR SST for NEA searches \citetext{T. Spahr, priv.\ comm.}. We note that while the discovery rate has increased in absolute number almost continuously since 1998, this increase has been due mainly to an increase in the discoveries of the smaller $H>22$ NEAs. Meanwhile, the discovery rate of $H\leq22$ NEAs has remained flat at $\sim$365/year thanks to a continuous decrease in the discovery of $H<18$ NEAs being offset by a similar increase in the discovery of $18\leq H<22$ NEAs (see Figure~\ref{disc_rate}). At this rate (and bearing in mind discovery rate will drop as the number of undiscovered objects decreases and if survey capabilities remain unchanged), the Congressional mandate of 2005 \citep{Brown2005} to find at least 90\% of the NEAs larger than 140~m \citep[$H\leq22$, an estimated $13,200\pm1,900$ according to][]{Mainzer2011b} will be met \textit{no earlier} than 2030, i.e., more than 10~years late. \cite{Mainzer2011b} also predict there are $20,500\pm2,000$ NEAs larger than 100~m ($H\lesssim23$).

While the pace of NEA discoveries has increased each year, characterization is falling farther behind.

\subsection{Sizes}
\label{sizes} 

Only a small fraction ($\sim$7\%) of NEAs have well-determined diameters and albedos from thermal IR detection with NEOWISE \citep{Mainzer2011b} and Spitzer \citep{Thomas2011} and fewer than 5\% have been detected by ground-based radar\footnote{\url{http://echo.jpl.nasa.gov/asteroids/index.html}, accessed 2014/05/01.}. While a higher radar rate ($\sim$100/year) has now been implemented and higher spatial resolution is being investigated, the total number of NEAs accessible to radar is limited (by the outgoing pulse power) to those that come well within $\sim$0.1~AU\footnote{See the SNR graphs for the Arecibo (\url{http://echo.jpl.nasa.gov/~lance/snr/far_asnr18.gif}) and Goldstone (\url{http://echo.jpl.nasa.gov/~lance/snr/far_gsnr-20.gif}) radar dishes}. Although $\sim$61\% of NEAs have $\rm MOID<0.1$~AU, the correct positioning to achieve a $<$0.1~AU distance from Earth occurs rarely. With the revived NEOWISE mission, the prospect of more thermal IR detections has improved. In 3~years $\sim$2,000 known NEOs (Near-Earth Objects) will be detected and then albedos measured to $\sim$50\%, for a total of $\sim$25\% of known NEOs\footnote{\url{http://www.lpi.usra.edu/sbag/meetings/jan2014/presentations/08_1415_Mainzer_SBAG.pdf}}.

For the vast majority of NEAs these quantities must be inferred from ground-based visible/near-IR observations yielding estimates of absolute magnitude $H$ and albedo. However, the $H$ alone is often uncertain by up to a magnitude due to inhomogeneous photometry and star catalog magnitude biases  \citep{Williams2015}, light curve variability \citep[amplitudes typically $\sim$0.4~mag,][]{Mainzer2011b}, poorly-constrained phase curves, and systematic size-dependent biases of unknown cause  \citep[e.g.,][]{Chesley2002,Pravec2012}.

\subsection{Rotation Periods}

By the beginning of 2014, $<$8\% of all known NEAs had period determinations listed in the Light Curve Database \citep{Warner2009}, with the number increasing at a rate of 150--200 NEAs per year \citetext{B. Warner, priv.\ comm.}, mostly for $V<18$ objects. We note that many of the light curves these rotation periods are derived from are not published (e.g., the MPC receives light curve data for only $\sim$30 NEAs per year\footnote{\url{http://www.minorplanetcenter.net/light_curve}}).

\subsection{Composition}

Spectral characterization of NEAs proceeds at a similarly slow rate. The largest program of optical-near-IR spectroscopy (0.5--2.5 microns) is the MIT-IRTF programme\footnote{\url{http://smass.mit.edu/minus.html}}, which has acquired a total of $\sim$1,000 spectra (of mostly large $H<15$ NEAs) at a rate of $\sim$100/year. It would take almost two centuries to complete spectroscopy of all $\sim$20,500 NEAs with $H<23$ and $\sim$120~years for the $\sim$13,200 NEAs with $H\leq22$ \citep{Mainzer2011b} at current rates, should these be the goals of such an IR spectroscopy program. It is noted, as we will show in \S\ref{subsequent}, that most of these NEAs will not be bright enough for ground-based IR spectroscopy.

\subsection{Orbital Uncertainties}

Orbital uncertainties are large (greater than 6~arcmin/decade along the orbit) for nearly all objects smaller than $H=22$ owing to a lack of long-arc astrometry. And, as we show in \S\ref{orbit_uncertainty}, astrometric follow-up is also lagging.

\subsection{Trends}

\begin{figure}[!ht]
\begin{center}
\includegraphics[width=0.99\linewidth]{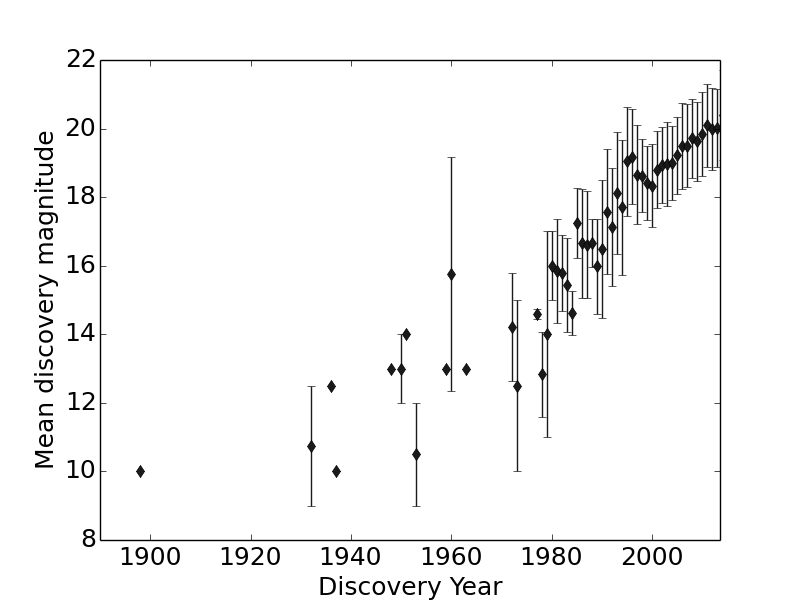}
\caption[Mean Discovery Magnitude]{Mean discovery magnitude for NEAs (approximately $V$ band, see \S\ref{discovery}). Bars represent 1-$\sigma$ scatter around the mean. There is a steady 1~mag increase in depth every $\sim$7.5~years visible in the past few decades.  The current mean is 20.0 with 1-$\sigma$ scatter of 1.0~mag.}
\label{meandiscoverymags}
\end{center}
\end{figure}

\begin{figure}[!ht]
\begin{center}
\includegraphics[width=0.99\linewidth]{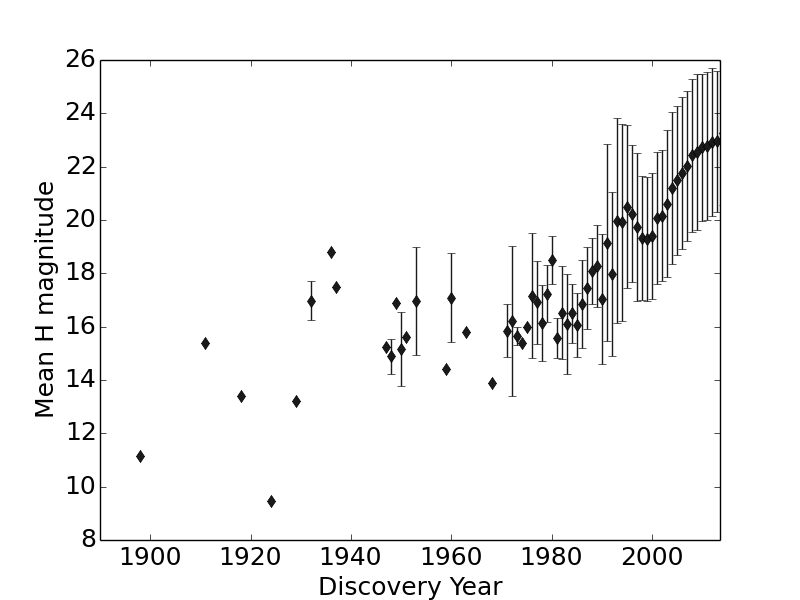}
\caption[Mean H Magnitude]{Mean $H$ magnitude of discovered NEAs by year. Bars represent 1-$\sigma$ scatter around the mean.}
\label{meanHmags}
\end{center}
\end{figure}

Follow-up observations will become even more challenging as surveys become sensitive to fainter objects and as large NEAs become fully known, as they effectively already are for objects with diameters $>$1~km \citep{Mainzer2011b}. The progression of NEA discovery magnitudes (see \S\ref{brightness}) is shown in Figures~\ref{meandiscoverymags} and \ref{meanHmags}.  For the past three decades there has been a steady increase in mean discovery depth by about 1~magnitude every $\sim$7.5~years, from $\langle V\rangle=16$ to $\langle V\rangle=20$. The mean $H$ magnitude has become correspondingly larger, from $\langle H\rangle\sim15$ in 1980 to $\langle H\rangle\sim22$ in 2013, corresponding to an approximate size decrease from 3,550~m to 140~m. 

\begin{figure}[!ht]
\begin{center}
\includegraphics[width=0.99\linewidth]{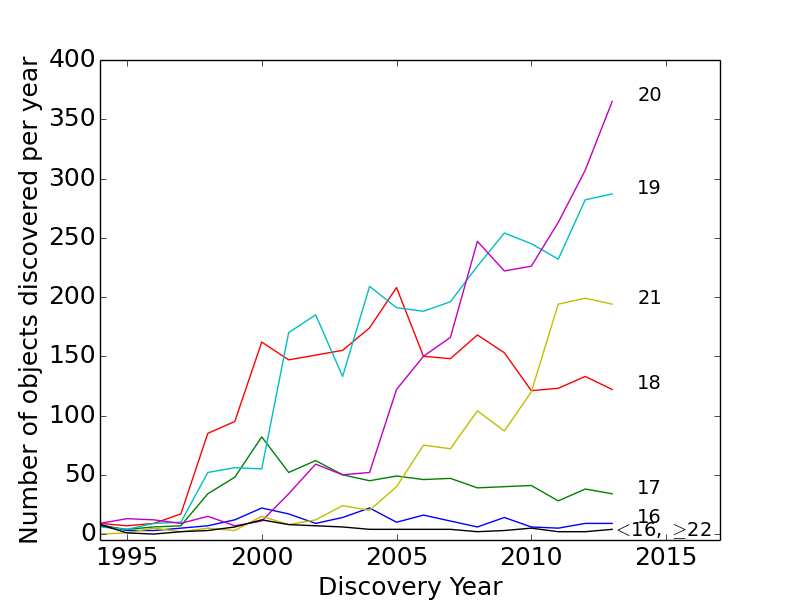}
\caption[Discovery history by magnitude]{Discovery magnitude (approximately $V$ band, see \S\ref{discovery}) for objects discovered since 1994.  Lines are labeled by magnitude $m$ representing the range $m\leq V_{\text{disc}} < m+1$.  The changing slopes indicate how the brighter discoveries have been saturated while the fainter ones are increasing. The annual counts are useful for planning follow-up programs.}
\label{dmaglines}
\end{center}
\end{figure}

Figure~\ref{dmaglines} shows in more detail the breakdown of discovery magnitudes over the past 20~years. The discovery rate at $V\leq18$ saturated at $\sim$200/year in 2000 as most of the big objects were found. Meanwhile, the discovery rate at $V=20\mbox{--}21$ has increased over the same period and now dominates at $\sim$375/year.

%%%%%%%%%%%%%%%%%%%%%%%%%%%%%%%
% RESULTS
%%%%%%%%%%%%%%%%%%%%%%%%%%%%%%%
\section{NEA Brightnesses}
\label{brightness}

%%%%%%%%%%%%%%%%%%%%%%%%%%%%%%%%%%%%%%%%

\subsection{Brightness at Discovery}
\label{discovery}

\begin{figure}[!ht]
\begin{center}
\includegraphics[width=0.99\linewidth]{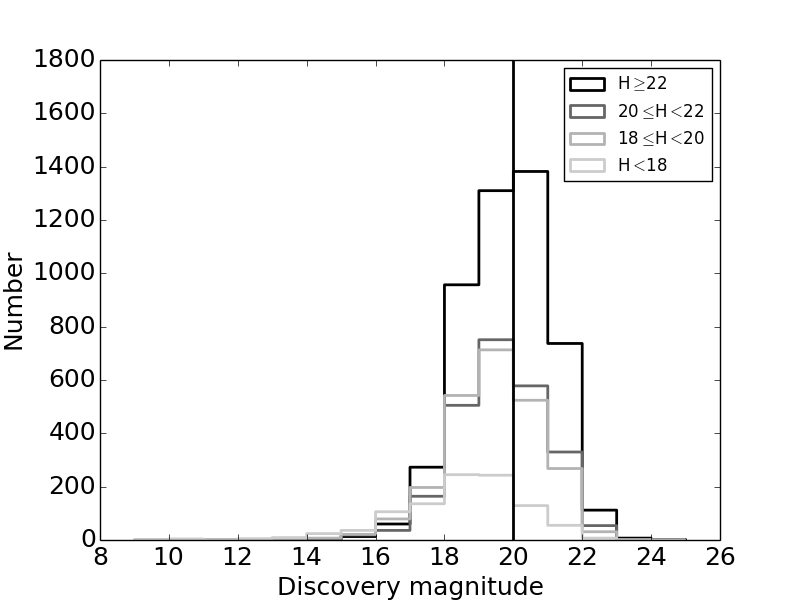}
\caption[Discovery Magnitude]{Discovery Magnitude. First recorded optical magnitude (which we use as a proxy for $V$; see text) of NEAs known as of March 2013 (9,723 objects). The current mean discovery magnitude, $V\sim20$, is shown for reference.}
\label{discV}
\end{center}
\end{figure}

To characterize the brightnesses of NEAs at discovery, we have combined the MPC Atens/Apollos/Amors files, extracted the discovery dates of the nearly 10,000 NEAs known as of 2013 March, and looked up their discovery observation optical magnitude through the NEODyS-2 interface.  These are usually $V$ or $R$ magnitudes, which we use with no correction given our coarse magnitude binning, but we note that we might expect $V-R$ as big as $\sim$0.4--0.5~mag based on Sloan Digital Sky Survey (SDSS) color-color plots for main belt objects in \cite{Ivezic2001} and transformation to $R$ and $V$ through the SDSS photometric relations\footnote{\url{http://www.sdss.org/dr7/algorithms/sdssUBVRITransform.html}}. We also note that Pan-STARRS discoveries have been reported in, e.g., the $r$, $i$, and wide $w$ bands, but the offsets from these to $V$ are expected to be 0.1--0.4~mag for most asteroids \citep{Denneau2013}. There are a few objects (currently $<$0.2\%) that were WISE discoveries with no magnitude for the discovery date. Otherwise we expect that our technique provides a reasonable approximation of $V$ band magnitude at discovery for nearly all the objects.  The discovery magnitude distribution is shown in Figure~\ref{discV} for various $H$ bins. The overall mean is 19.4 with a standard deviation of 1.4.

\begin{figure}[!ht]
\begin{center}
\includegraphics[width=0.9\linewidth]{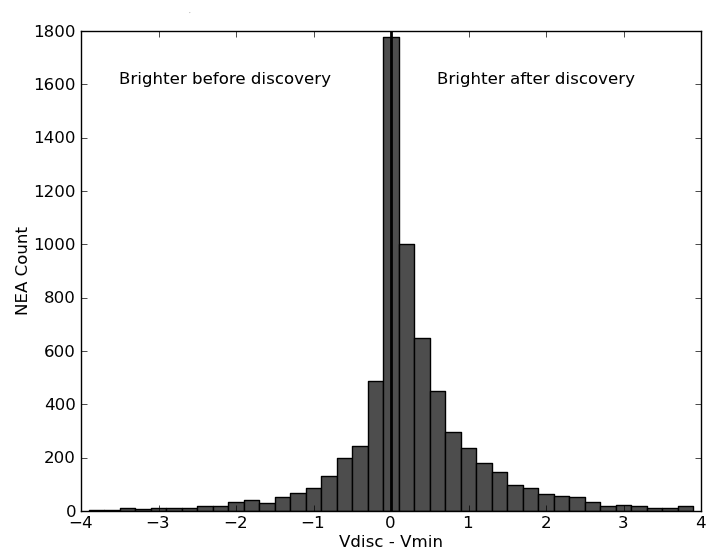}
\caption[Difference between discovery and brightest magnitude]{Distribution of the difference between discovery and brightest magnitude, $V_{\textrm{diff}}$, for the 6,763 NEAs in the Decade Sample. NEAs discovered before reaching peak brightness have positive values of $V_{\textrm{diff}}$, while those discovered after have negative values. 57 NEAs with $|V_{\textrm{diff}}|>4$ were omitted for clarity.}
\label{Vchange}
\end{center}
\end{figure}

To explore how a typical NEA's brightness changes during its discovery apparition, we consider the Decade Sample.  For each NEA in this sample, we look up the discovery date in the Atens, Apollos, and Amors data files, and then examine the ephemerides at daily intervals to determine how the asteroid's magnitude (theoretically) changed throughout its discovery apparition. We define 
$V_{\textrm{diff}} = \pm |V_{\textrm{disc}} - V_{\textrm{min}}|$, where $V_{\textrm{disc}}$ is the discovery magnitude and $V_{\textrm{min}}$ is the brightest magnitude achieved during the apparition, and where +/- values represent NEAs brighter after/before discovery, respectively. 
Figure~\ref{Vchange} gives the distribution of $V_{\textrm{diff}}$.  NEAs discovered before reaching peak brightness have positive values of $V_{\textrm{diff}}$, while those discovered after have negative values. Most NEAs ($\sim$60\%) are discovered within 0.5~magnitudes of their peak brightness during the discovery apparition.

\begin{figure}[!ht]
\begin{center}
\includegraphics[width=0.99\linewidth]{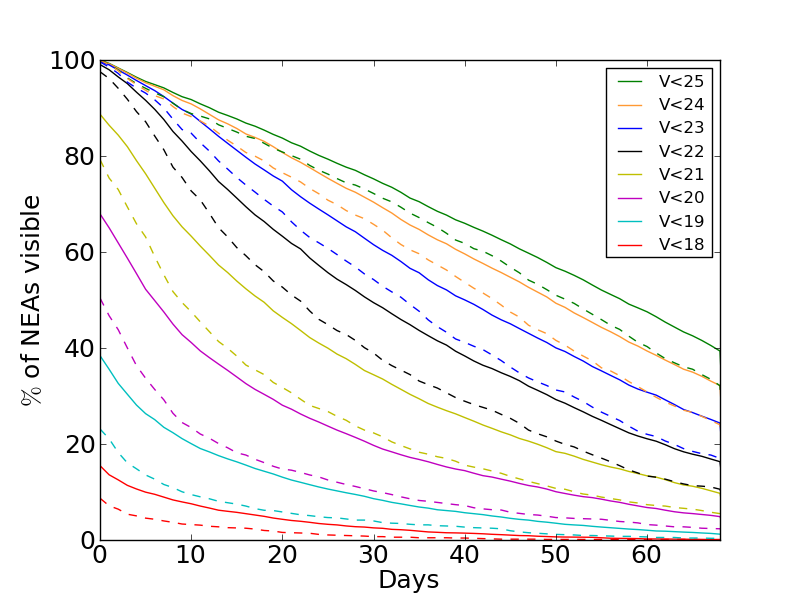}
\caption[Observable days following discovery]{Observable days following discovery for the 6,763 NEAs in the Decade Sample. Percentage of NEAs that remained brighter than $V = 18$, 19, 20, 21, 22, 23, 24 and 25 for up to 10~weeks following their discovery. Solid lines are for the full Decade Sample; dashed lines are for the last 2~years of the sample only.}
\label{days_bright}
\end{center}
\end{figure}

We then evaluated the fraction of NEAs remaining brighter than a series of limiting apparent magnitudes, $V_{\textrm{lim}}$, after discovery (Figure~\ref{days_bright}). It is clear that this fraction drops rapidly for many interesting cases. The values of $V_{\textrm{lim}}=18$ was motivated by the infrared spectroscopic capabilities of the IRTF, and also is in range for photometric and astrometric follow-up by amateurs. $V_{\textrm{lim}}=20$ represents the infrared spectroscopic limit of Magellan (see \S\ref{discussion}) and is the current mean discovery magnitude. $V_{\textrm{lim}}=21$ represents the limit for the proposed NEA spectroscopic follow-up project LINNEAUS \citep{Elvis2014c}. The solid lines show the brightness evolution using the full Decade Sample, while the dashed lines are drawn from only the NEAs discovered during the last 2~years of the Decade Sample. As can be seen, the number of NEAs that remained above a certain brightness decreased noticeably over time, due mainly to the increase in the number of small ($H>22$) NEAs discovered (Figure~\ref{disc_rate}) and the greater overall survey depth in more recent years (Figure~\ref{meandiscoverymags}).

Approximately 2/3 of the objects discovered during the decade were found at $V\leq20$, but faded by 0.5~mag after $\sim$1~week, 3.5~mag by $\sim$4 weeks, and by more than 5~mag by $\sim$6~weeks. Meanwhile, at the bright end, we see that only $\sim$17\% of NEAs became brighter than $V=18$, even during their discovery apparition (Figure~\ref{days_bright}), and even less, $\sim$3\%, remained brighter than $V=18$ for more than 30~days. The small objects ($H\gg22$) that are only seen on very close passages would fade much faster than these aggregate values, as evidenced by the dashed lines, which show the values of these curves for the last 2~years of the Decade Sample. These last 2~years featured many more small ($H>22$) NEA discoveries than for the average of the sample. The discovery apparition then offers only a short window for characterization. We explore the implications for follow-up in \S\ref{discussion}.  

\subsection{Brightnesses on Subsequent Apparitions}
\label{subsequent}

NEAs will have other apparitions following discovery, so we also investigated their observability at these subsequent apparitions. The success of a ``subsequent apparition'' approach to characterization depends on two factors:

\begin{enumerate}
\item The length of time that will elapse before the NEA's next apparition.
\item The brightness the NEA will reach on its next apparition(s).
\end{enumerate}

The synodic period, $S$, is the time that elapses between two consecutive apparitions (i.e., when the NEA returns to a similar position near Earth). If an NEA's orbital period is denoted by $P$ (given in years), then its synodic period with respect to Earth will be given by \citep{Universe}: 

\begin{equation}
S = \bigg|1 - \dfrac{1}{P}\bigg|^{-1} \text{ years}
\end{equation}

\begin{figure}[!ht]
\begin{center}
\includegraphics[width=0.99\linewidth]{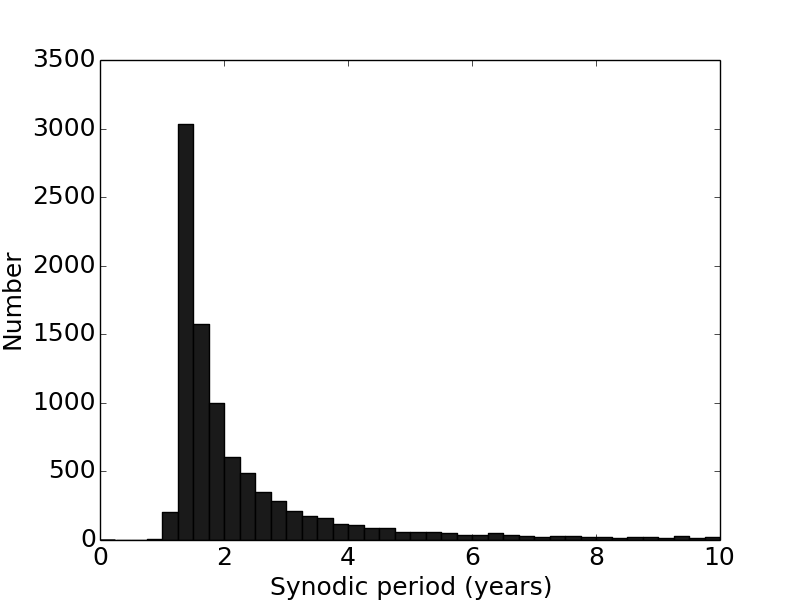}
\caption[Synodic Period]{Synodic Period distribution of NEAs known as of March 2013 (9,723 objects). For clarity, the 5\% of outliers with $S>10$~years are not plotted.}
\label{synodic_periods}
\end{center}
\end{figure}

The distribution of synodic periods for the known NEAs as of March 2013 is shown in Figure~\ref{synodic_periods}.

Almost 80\% of known NEAs have $S<3$~years and will therefore have at least three apparitions each decade. Synodic periods of 3--10~years account for $\sim$20\% of known NEAs. These reapparition times are too long for space mission target selection studies as the properties and highly accurate orbits needed for selection will take too long for planning purposes.

As an example of what it takes to determine an orbit well enough for space mission planning, we present NEA (101955) Bennu, the target for the OSIRIS-REx mission\footnote{\url{http://www.asteroidmission.org/}}. To obtain an accurate enough ephemeris, it was observed with radar during 3 apparitions and had optical observations spanning 15 years. This provided a very accurate orbit determination, which also constrained the magnitude of the Yarkovsky effect acting upon it, thus allowing the mass/area ratio of Bennu to be calculated \citep{Chesley2014}. 

Only 5\% of the nearly 10,000 NEAs known as of 2013 March have $S>10$~years as their position relative to Earth changes very slowly; it will be decades before they can be re-observed from Earth after their discovery apparition. There may also well be NEAs with long synodic periods currently ``hiding'' within the region of space at elongations $<$60\degree, making them unobservable because of their position with respect to the Sun; these may not be discovered from Earth for decades. The positive side of this situation is that once a long synodic period NEA becomes observable, it will remain so for an extended period of time. 

\begin{figure*}[!ht]
\begin{center}
\includegraphics[width=0.99\linewidth]{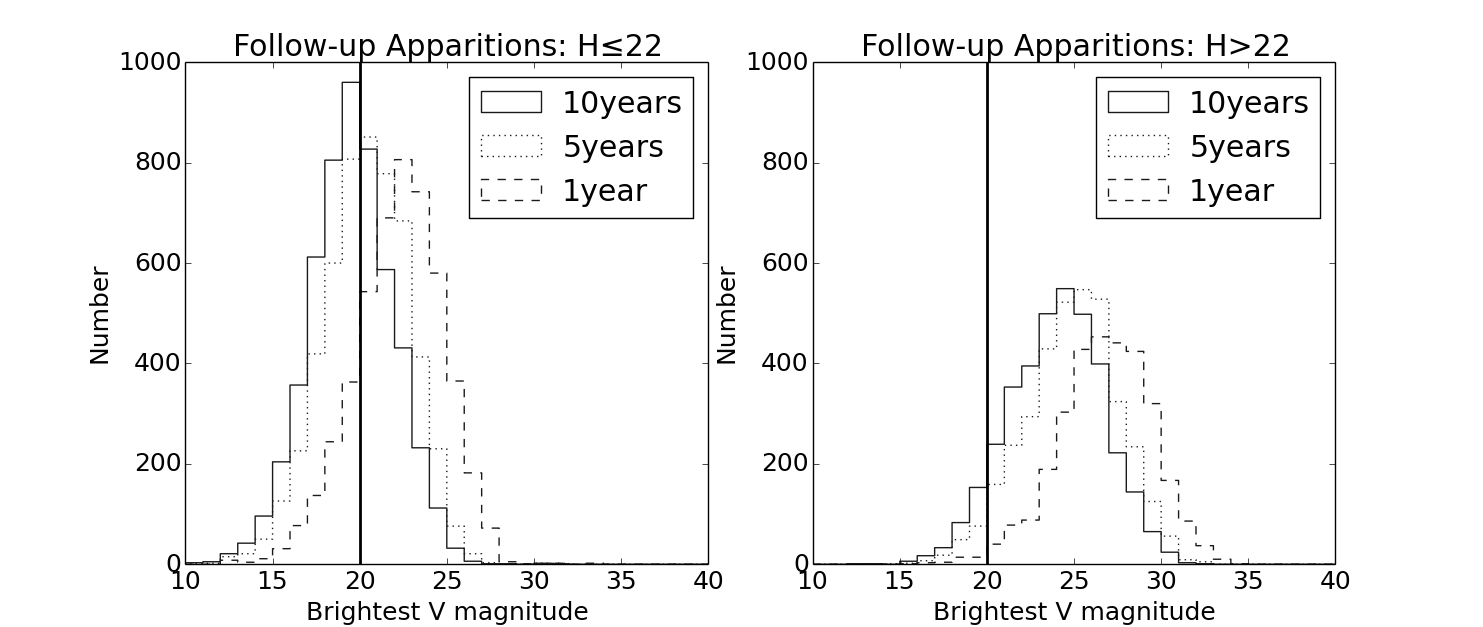}
\caption{Follow-up apparitions of known (as of March 2013) NEAs over a 1, 5 and 10 year period (starting January 1, 2018) for (left) $H\leq22$, and (right) $H>22$, showing the distribution of the brightest $V$ magnitude that will be reached.  The current mean discovery magnitude, V$\sim$20, is shown for reference. Note that fainter objects are to the right in each plot.}
\label{follow_up}
\end{center}
\end{figure*}

Given that most of objects will return several times within a decade, we computed the brightest magnitude that NEAs known as of 2013 March will achieve on return apparitions over a 10 year period starting January 1, 2018, by which time, most of today's newly discovered NEAs would be making at least their first return.  We used the MPC tool \texttt{MPEph} (see \S\ref{data_sources}) to generate daily ephemerides for these NEAs and determined the brightest $V$ magnitude reached over a 1, 5, and 10~year period (starting January 1, 2018). Results are shown in Figure~\ref{follow_up} (on the left for objects with $H\leq22$ and on the right for $H>22$). 

Figure~\ref{follow_up} shows that over a period of one year, less than 1\% of the known NEA population will become brighter than $V=18$. Even over a decade, only $\sim$17\% will reach $V<18$. By comparison to Figure~\ref{discV}, we see the larger NEAs ($H\leq22$) typically do return to similar brightnesses within a decade; however, smaller NEAs ($H>22$) are generally no brighter than 5~magnitudes fainter than at discovery, and rarely brighter than $V=21$. This reflects the bias towards discovering small NEAs during their very closest approaches to Earth.

Most of the NEAs that remain to be discovered and characterized are at the smaller end of the size distribution\footnote{\url{http://sservi.nasa.gov/wp-content/uploads/2014/03/Harris.pdf}}; if follow-up of these objects is not performed during the discovery apparition, the distribution in Figure~\ref{follow_up} suggests that nearly all will remain uncharacterized for over a decade. We conclude that from a brightness point-of-view, follow-up cannot wait for a subsequent apparition but must be performed during the discovery apparition.

\section{NEA Positional Constraints}
\label{positional_constraints}

\subsection{Orbit Uncertainty}
\label{orbit_uncertainty}

Following discovery, an NEA's orbit must be determined to sufficient accuracy for the NEA to be successfully re-acquired for follow-up on a subsequent apparition. The uncertainty in an NEA's orbital solution is quantified by the MPC via the Orbit Uncertainty Parameter, $U$\footnote{\url{http://www.minorplanetcenter.net/iau/info/UValue.html}}. $U$ is an integer running from 0 (extremely precise orbit) to 9 (highly imprecise orbit). An asteroid with a $U$ value of 0 is defined to have an orbit precise enough to allow calculation of its position along its orbital path to within $\pm1''$ after one decade. Each integer step in $U$ is a logarithmic step increase of a factor $\sim$4.4 in this uncertainty. So $U=9$ corresponds to an uncertainty of more than $\sim$41\degree\ after a decade. It is important to understand that $U$ is associated with an uncertainty in position \textit{along the orbit} of the asteroid, which cannot be readily converted to an on-sky positional uncertainty because it would depend on the asteroid's trajectory with respect to the Earth observer at the time of the observation.

With sufficient observing time and telescope aperture, any NEA of sufficient brightness can be recovered at subsequent apparitions after discovery, but practically speaking, to lie within a typically-sized imaging field ($\sim$10~arcmin), requires $U\leq3$ (for an NEA; for Main Belt Asteroids the tolerances are more relaxed). This quality of $U$ has to be reached during the discovery apparition to allow for ready recovery, and thus further orbit quality improvement at subsequent apparitions. Radar has the potential to greatly improve the quality of a calculated orbit given that a radar observation can reduce on-sky positional uncertainty of an NEA at the following apparition by up to 2 orders of magnitude or more \citep{Ostro2002,Ostro2007}. Unfortunately, only $\sim$100 NEAs are observed each year with radar and, as we discuss in \S \ref{sizes}, many NEAs are out of reach of current radars.

\begin{figure}[!ht]
\begin{center}
\includegraphics[width=0.99\linewidth]{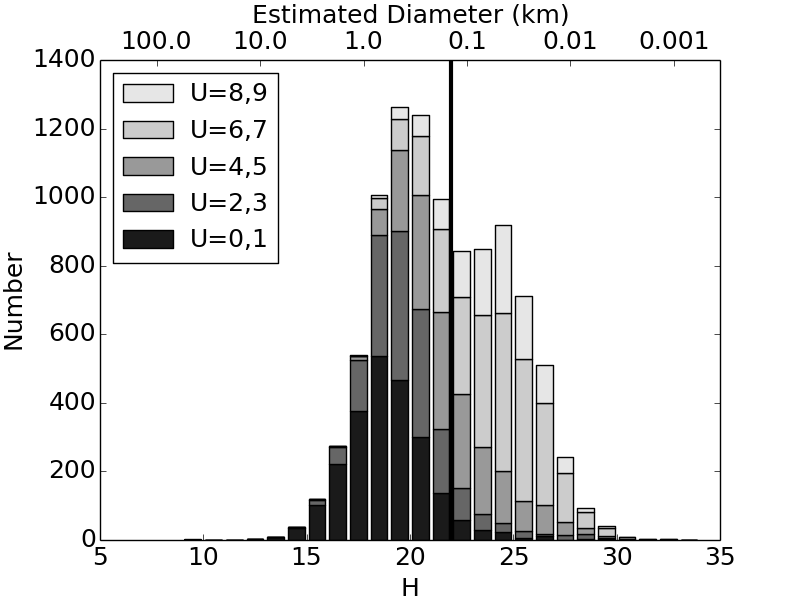}
\caption[Orbit Uncertainty vs. H magnitude]{Orbit Uncertainty vs. $H$ magnitude for NEAs discovered before 2013 March. A line marking $H=22$ is plotted for convenience.}
\label{fig_orbit_uncertainty}
\end{center}
\end{figure}

We used the MPC \texttt{NEA.txt} to extract the values of $U$, $H$, number of oppositions, and arc length (given as days for single opposition objects, or years between first and last observation for multi-opposition objects). Figure~\ref{fig_orbit_uncertainty} shows the $U$ values as a function of the $H$ magnitude for NEAs discovered through 2013 March\footnote{The estimated diameter was calculated using the following relationship \citep{FowlerChillemi1992} between the diameter, $D$, visual albedo, $p$, and absolute magnitude, $H$: 
\begin{equation}
D = \frac{1329}{\sqrt{p}}10^{-0.2H} \text{ kilometers}
\end{equation}We used a value of $p=0.14$, as recommended by \cite{Stuart2004}.}. Over 65\% of larger ($H\leq22$) NEAs have values of $U\leq3$, and thus, orbits well-defined enough for re-acquisition on subsequent apparitions. 
These high-quality orbit determinations come about for two reasons: (1) many were discovered over a decade ago so they have been observed over several apparitions; and (2) being large, they are generally brighter and thus can be discovered (or reobserved) farther from Earth, where their relative motion to Earth leads to slower fading than is the case for smaller objects traveling closer. 

However, for smaller NEAs ($H>22$) less than $10\%$ have $U\leq3$. By $H>25$ this fraction is reduced to $\sim$5\%.  Only more recent surveys have been able to find significant numbers of $H>22$ NEAs (Figure~\ref{disc_rate}).  In addition, this population tends to return at fainter magnitudes (see \S\ref{subsequent} and Figure~\ref{follow_up}). Therefore, few have been observed over multiple apparitions.  The overwhelming fraction of known $H>22$ are thus poor candidates for targeted re-acquisition on the next apparition and will have to be re-discovered by chance. There are only $\sim$50 very small NEAs ($H>25$) with very well-determined orbits ($U\leq2$) such that they could be found beyond the next apparition with a targeted observation.

At $H>26$ the fraction of NEAs with smaller $U$ shows a small increase. This we attribute to the fact that smaller NEAs must be physically close to Earth to become bright enough to be detected. When an NEA is discovered close to Earth, its initial orbit will not generally be accurate enough to entirely rule out an impact with our planet. Possible impactors are heavily observed until an improved and more accurate orbit is obtained, with a sufficiently low value of $U$ to rule out an impact with 100\% certainty.

\subsection{Arc Length}
\label{arc_length}

The most direct way to reduce the uncertainty in an NEA orbit is to increase the arc length observed during the discovery apparition, and then through re-observation at subsequent apparitions. An NEA's arc length is simply the number of days elapsed between the first and last time it was observed.

\begin{figure}[!ht]
\begin{center}
\includegraphics[width=0.99\linewidth]{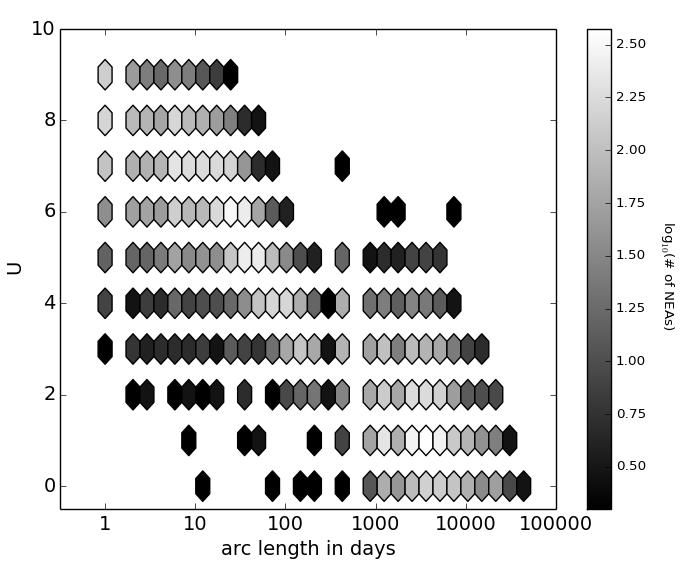}
\caption[U vs. arc length]{$U$ vs. arc length for NEAs discovered as of 2013 March. Data point density is represented by the color scale.  Arc lengths for multiple apparition objects are plotted as 365 days times the number of years between first and last observation.  The gap at 1 year marks the boundary between single opposition and multiple apparition objects.}
\label{UvArc}
\end{center}
\end{figure}

A clear relation of $U$ with $\log_{10}(\text{arc length})$ in days is shown in Figure~\ref{UvArc}. This figure can be compared to Figure~4 in \cite{Desmars2013}, who carried out a similar exercise using the current ephemeris uncertainty as a measure of the orbital uncertainty for all known asteroids (not just NEAs). They found that astrometry conducted within the first 2~days after discovery had the highest impact in reducing positional uncertainty for NEAs; astrometry from that point on until 7--14 days had little effect on the uncertainty, after which point the uncertainty begins to decrease with increasing arc length. Our results corroborate this for the NEAs: Up to $\sim$10~days, the $U$ is typically $\sim$7. \cite{Desmars2013} also found that astrometry on 10--250 day timescales improves the orbit much more rapidly. Our plot bears these results out for NEAs as well: $U$ decreases linearly as a function of $\log_{10}(\text{arc length})$ for arc lengths $\gtrsim$10~days. There is a clear ridge line in our figure that defines the typical arc length needed to reach a given $U$ value, and by 100--200 days the $U$ has typically dropped to $\sim$3. The relation between $U$ and arc length extends to multiple apparitions. However, even with multiple apparition observations, about $\sim$40\% of NEAs still have $U\ge2$; this is generally due to an NEA having been observed for short arcs during a few (often non consecutive) apparitions, stressing the need for longer individual observing arcs during each apparition. 

Achieving $U\leq3$ at the discovery apparition typically requires astrometry over an arc length of $\gtrsim$3~months (Figure~\ref{UvArc}). For $H\leq22$ NEAs, obtaining such a long arc requires large telescopes (4~m class or larger) to ensure the object is detected (see \S\ref{discussion}).  However, the extent to which further astrometric measurements reduce $U$ for specific NEAs will depend on the phase of the orbit being observed, as determining orbit curvature is crucial. We will explore this in a subsequent paper. We did search for differences in the arc length vs $U$ relation for Aten, Apollo and Amor NEAs separately, but found no difference.

\section{Sky Motion}
\label{sky_motion}

\begin{figure}[!ht]
\begin{center}
\includegraphics[width=0.9\linewidth]{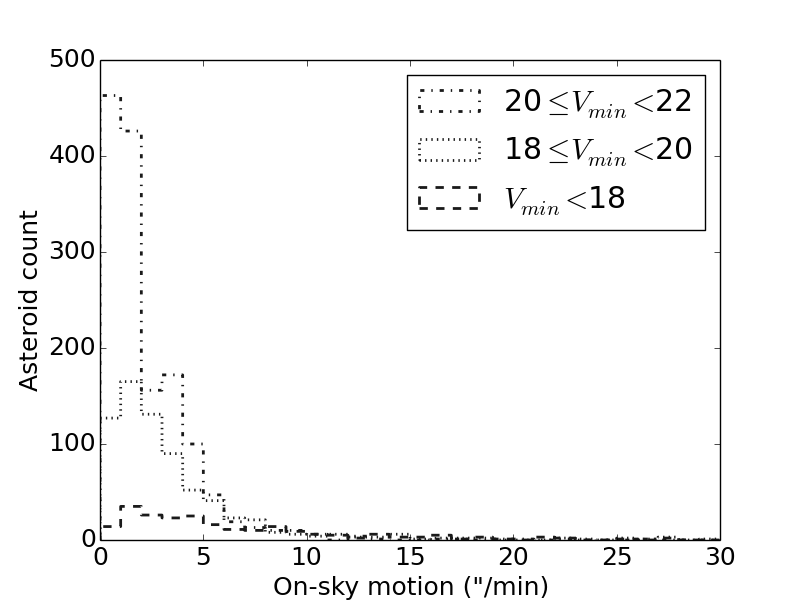}
\caption[Sky Motion]{Sky Motion of NEAs visible during 2013 (with elongations $>$60\degree). For each NEA, the sky motion on the night of maximum brightness was used. Outliers with motions $>$30$''$/min ($\sim$1\% of the sample) are not shown.}
\label{motion}
\end{center}
\end{figure}

The relative proximity of NEAs means they have proper motions that can potentially limit follow-up observations.  In this section we explore NEA sky motions during a typical year. Using \texttt{NEAobs} (see \S\ref{data_sources}) we created ephemerides for all NEAs visible during 2013, and determined the sky motion on the night when each NEA was at its brightest. Figure~\ref{motion} shows the resulting distribution.

We find that the typical sky motions are $<$5$''$/min, with median $\sim$1.5$''$/min. This plot can be used to estimate the fraction of targets accessible to a telescope conducting follow-up observations from a sky motion perspective.  For example, less than 5\% of NEAs visible in 2013 were moving faster than the 9$''$/min that the Magellan telescope is capable of tracking, so sky motions would not pose a significant limitation for this case.  On the other hand, for any follow-up observations done by sidereal tracking, the sky motion constrains exposure times as described in \S\ref{discussion}.

\section{Hemisphere Bias}
\label{hemisphere_bias}

All large NEA ground-based surveys, except the recently cancelled Siding Spring Survey in Australia, are in the Northern Hemisphere, so $\sim$30\% of the most southern part of the celestial sphere is not searched as thoroughly as the rest.  Our ephemerides show that 92\% of the known NEAs will appear in both hemispheres within a five year period, and 98\% within a ten year period.  For $H\leq22$, this may permit follow-up from either hemisphere on subsequent apparitions.  However, given that for $H>22$ the objects will return $\sim$5~mag fainter (see \S\ref{subsequent}, Figure~\ref{follow_up}), follow-up may not be possible from \textit{either} hemisphere. This is investigated further in a future paper.

The limitations discussed in \S\ref{discovery} (see Figure~\ref{follow_up}) highlight the need to complete most of the follow-up at discovery. For this reason, some follow-up facilities need to be placed at similar locations to the telescopes being used for discovery. As of 2012, $\sim$67\% of NEAs are being discovered from the U.S. Southwest region (in an area around $\sim$32\degree N $\sim$110\degree W) by the Catalina Sky Survey and associated Mt. Lemmon Survey, LINEAR\footnote{Lincoln Lab's Near Earth Asteroid Research program, which was terminated as of 2013.}, and Spacewatch. Rapid (discovery night) follow-up would argue for very close proximity from a common-weather point of view, while longer-term follow-up (e.g., long-duration coverage of light curves or follow-up in subsequent nights after discovery) could benefit from observatories distributed along longitude and at similar latitudes. Follow-up of discoveries from Hawaii (where Pan-STARRS is currently the second most prolific NEA discovery site) must be made in Hawaii, as the next high-quality observing site to the West is in the Canary Islands, some 15~h away. If rapid follow-up is desired, then this also argues for astrometry of discovery observations to be submitted to the MPC in near realtime.

\section{Discussion}
\label{discussion}

Our analysis of the currently-known population of NEAs shows that:

\begin{enumerate}[1)]

\item{ The discovery rate of $H\leq22$ NEAs has remained steady at $\sim$365/year over the past decade (see Figure~\ref{disc_rate}), suggesting it will take over 20~years to discover the remaining population of $H\leq22$ NEAs unless the rate of $18\leq H<22$ NEA discovery increases dramatically.}

\item{ The mean magnitude for NEAs being discovered today is $H\sim23$, which has been increasing at a rate of $\sim$1~mag every 3~years. Meanwhile, the mean discovery magnitude for NEAs is $V\sim20$, which has increased at a rate of 1~mag every $\sim$7.5~years.}

\item{ $\sim$60\% are discovered within 0.5~mag of their peak brightness during the discovery apparition and they fade quickly: the typical NEO will dim by 0.5~mag after $\sim$1~week, 3.5~mag by $\sim$4 weeks, and by more than 5~mag by $\sim$6~weeks.}

\item{ 80\% of synodic periods (which define windows for followup in subsequent apparitions) are $<$3~years so that 3 apparitions per decade are available for charactrization if the returning NEA is sufficiently bright.}

\item{ The average brightness for $H>22$ NEAs on return apparitions in the decade following their discovery is 5~magnitudes (i.e., 100 times) fainter than at discovery. For larger ($H\leq22$) NEAs, about half of the decade-scale returns achieve similar brightness to that of their discovery apparition.}

\item{ Orbital uncertainties are significant for the bulk of currently known NEAs, as most have been observed for relatively short arcs on a single apparition.  Whereas most (60\%) of $H\leq22$ NEAs have $U\leq3$, less than 10\% of $H>22$ NEAs do. To reduce the orbital uncertainty parameter on the discovery apparition to levels practical for recovery on subsequent apparitions ($U\leq3$) requires astrometric measurements over arcs $\gtrsim$ 3~months.}

\item{ Sky motions when NEAs are at their observable brightest are generally $<$5$''$/min, with median $\sim$1.5$''$/min except for very close approaches for which the rate can exceed this by an order of magnitude.}

\item{ Currently, the concentration of surveys in the Northern Hemisphere means that follow-up observatories must be located in the North too for observations on the discovery apparition. However,  follow-up of $H\leq22$ objects on subsequent apparitions can be done from North or South (assuming they can be found), as $>$92\% of these bigger NEAs will appear in both hemispheres within a 5 year period at magnitudes comparable to those at discovery.}

\end{enumerate}

We can use these results, along with projected discovery circumstances for the next few years, to determine telescope requirements for various types of follow-up observations.  

\subsection{Fiducial Magnitudes}
\label{fiducial_magnitudes} 

We begin by defining a few fiducial magnitudes for exposure time predictions. The mean magnitude at discovery is currently $V=20.0\pm1.0$~mag (see Figure~\ref{meandiscoverymags}), so to reach the bulk ($\sim$2/3) of these objects at discovery, observers would need to achieve depths of $V\sim21$.  However, the mean discovery magnitude among all surveys increases by $\sim$1~mag every 7~years, and Pan-STARRS is already achieving this today.  Thus, we take $V\sim21$ as the projected mean discovery magnitude for a year or two from now (and we should expect that to reach $\sim$2/3 of the population we would have to get a magnitude or so fainter, to objects with discovery magnitudes of $V\sim22$).  

In the discussion that follows, we take the historically-preferred $V$ magnitudes and translate them to $R$ (or AB mag $r$), which are more likely representative for the follow-up observations. We do so crudely by taking $V-R\simeq0.5$ and note that $r-R\simeq0.2$ (AB mag to Vega mag, see section \S\ref{brightness} above). With these assumptions and our results from above, we derive the following fiducials:

\begin{itemize}
\item {\it The fast fading of newly-discovered asteroids and 100$\times$ fainter reapparitions will require some types of follow-up to be done rapidly (within days) after discovery.}  Taking the projected mean discovery magnitude of $V=21\pm1$, ($R=20.5\pm1$) and allowing for 0.5~mag of fading in a week, we find that the observations will need to achieve mean depths of $V\sim21.5$ ($R\sim21$), or for bulk of new objects $R\sim22$.  
\item {\it For follow-up observations done on a $\sim$months timescale after discovery} the NEAs are expected to be $>$3~mag fainter still; i.e., we require mean depths of $R\sim24$, or for the bulk of new objects $R\sim25$.
\item {\it For currently-known NEAs making return apparitions in the next decade}, we take the mean discovery magnitude for all NEAs discovered through 2013 March of $V\sim19.4$, with $V\sim21$ encompassing $>$2/3 of the objects (see \S\ref{discovery}).  For $H\leq22$ objects, the reapparitions will be similarly bright, setting the observing depths at $V\sim21$, or $R\sim20.5$ for the bulk of old objects.  However, for the $H>22$ objects the depths required to reach the mean is 5~magnitudes fainter.  That would be $R=25.5$ for the bulk of old objects, or $R\sim24.5$ for the bright half of old objects.
\end{itemize}

We show in Table~\ref{exp_time} the signal-to-noise (S/N) and exposure times for magnitude ranges around our fiducial magnitudes under dark-sky conditions typical of moonless nights on Kitt Peak and Cerro Tololo\footnote{\url{https://www.noao.edu/kpno/manuals/dim/}, accessed 2014/03/17.}. We have assumed the performance of a typical modern back-illuminated CCD and take the radius for photometry to be 2~arcsec sampled with sub-arcsecond pixels. (Thus for excellent seeing our estimates would be pessimistic.) Our calculations check out against the online exposure time calculators\footnote{\url{http://www.noao.edu/noao/noaonews/mar99/node16.html}}; e.g., the Kitt Peak 2.1~m and 4~m telescopes with these assumptions.

\begin{table*}[!ht]
\centering
\caption{S/N and exposure times for 2~m and 4~m telescopes}
\begin{tabular}{lccccccp{4.5cm}}
\toprule
\multicolumn{1}{c}{\multirow{2}{*}{R mag}} & \multicolumn{2}{c}{$\textrm{S/N}=4$} & \multicolumn{2}{c}{$\textrm{S/N}=10$} & \multicolumn{2}{c}{$\textrm{S/N}=100$} & \multicolumn{1}{c}{\multirow{2}{*}{Circumstances}} \\ \cmidrule(lr){2-7}
\multicolumn{1}{c}{}                       & 2 m         & 4 m         & 2 m          & 4 m         & 2 m          & 4 m          & \multicolumn{1}{c}{}                                                   \\
\midrule
20.5                                       & 16 s        & 2 s         & 50 s         & 10 s        & 45 min       & 10 min       & {\scriptsize Mean for new discoveries on same nights; or, most $H<22$ reapparitions} \\
22.0                                       & 90 s        & 16 s        & 7 min        & 90 s        & ...          & 2.5 h        & {\scriptsize Most new discoveries at $\sim$1~week} \\
24.0                                       & 35 m        & 10 m        & 4 h          & 1 h         & ...          & ...          & {\scriptsize Mean for new discoveries at $\sim$1~month} \\
25.0                                       & 4 h         & 1 h         & ...          & 5 h         & ...          & ...          & {\scriptsize Most new discoveries at $\sim$1~month; or, most $H>22$ reapparitions} \\                      
\bottomrule
\end{tabular}
\label{exp_time}
\end{table*}

\subsection{Requirements for Astrometry}
\label{astrometry} 

For ground-based astrometry we compute the exposure times needed to achieve tolerances of 0.1~arcsec. For faint objects, with a well-sampled PSF and no image trailing, astrometric precision should be given by \cite{Birney2006}:

\begin{equation}
\sigma_{ast} = \frac{1}{2.355} \times \frac{\text{FWHM}}{\text{(S/N)}}
\end{equation}

In 1~arcsec seeing we then require $\textrm{S/N}\sim4$ to achieve 0.1~arcsec precision, but as a source at that level is barely significantly detected, in practice one will likely require a slightly higher S/N.

The requirement for no NEA trailing could be satisfied in several ways.  One could track non-sidereally on the (potentially imprecisely known) motion of the asteroid, though centroiding the resulting star trails could introduce subtle astrometric biases, for example through timing errors caused by seeing/transparency variations during the exposure. Alternatively, one could take sets of individual exposures short enough to prevent trailing and then combine them later, either by shift-and-add to optimize S/N (which the MPC sternly warns against), or by averaging the individual position measurements. For the typical NEA motion of 1.5~arcsec/min found above, sidereally-tracked exposures would be limited to $t\sim40$~s.  

From Table~\ref{exp_time} we see that a 2~m telescope could accomplish the astrometric measurements even with sidereal tracking for newly-discovered objects on timescales of about 1~week (not surprising since the discoveries will be made by 2~m-class telescopes), and for reapparitions of known large NEAs (which will likely be done \textit{de facto} by future search surveys).  Long-arc (months) astrometry of only the brightest new objects could possibly be done on a 2~m by tracking on the asteroid and trailing the stars or shift-and-add. The mean new objects would require a 4~m telescope. Long-arc astrometry of the faintest new objects and the $H>22$ reapparitions would require a larger telescope still. {\it This argues strongly for conducting long-arc astrometry on 4+~m telescopes during the discovery apparitions.} The field of view would ideally be $\sim$20 arcmin to ensure that enough astrometric standards lie within each image and to accommodate uncertain positions. To ensure a well-sampled PSF would require pixel sizes of $\lesssim$0.5~arcsec.

\subsection{Requirements for Light Curves}
\label{light_curves}

For light curves, the S/N requirements are much stiffer. In addition, the temporal sampling defines the range of rotation rates for which periods can be determined, with one benchmark being that for ``sparsely-sampled" observations over many rotations that will undergo Fourier analysis, the light curve needs to be sampled at intervals no more than 18\% of the rotation period \citep{Pravec2000a}. The nearly 700 NEAs in the Light Curve Database \citep[see][]{Pravec2000b,Warner2009} with light curve properties listed show amplitudes of $\sim$0.4~mag, with rotation periods typically $<$2.2~h for $H>22$ objects (1/4 of the sample) and $>$2~h for larger ones.

However, the true distributions are not known, as NEAs with hard-to-measure periods would not be included in the database.  In their attempt to measure light curves of 83 sub-km NEAs with a 2.4~m telescope (down to $R\sim19$), \cite{Statler2013} found that 2/3 had ambiguous light curves that would not permit determination of the period. Their cadence was $\sim$90~s, and their median photometric precision was 0.04~mag. While their ambiguous cases were measured with typically worse than average precision, they concluded that the ambiguous objects were likely either super-fast or super-slow rotators, or objects with intrinsically low amplitude (e.g., axis-symmetric) objects. 

A recent detailed study by \cite{Warner2011b} finds that the predicted yield of Fourier analysis on sparsely-sampled (up to 16 times per night) light curves with noise of $\sim$0.03~mag is highest in the period range of 2--6~h and for amplitudes $>$0.2~mag. Such a strategy could be well-suited to reapparitions of the larger $H\leq22$ NEAs, typically exhibiting $>$2~h spin periods and amplitudes of $\sim$0.4~mag. For a 2~h period the 18\% requirement yields spacings of $>$20~min. Table~\ref{exp_time} shows that observations with a 1-sigma precision per point of a few percent, i.e., $\sim$0.03~mag \citep[yielding light curves with S/N $\sim$10 and corresponding to the cases discussed by][]{Warner2011b}, could be obtained with long blocks of time on 2~m-class telescopes. Wide-field surveys for transients are already providing periods for (mostly Main Belt) asteroids this way \citep[e.g.,][]{Chang2014}.

However, as \cite{Warner2011b} articulate, there are compelling reasons to measure the spin properties of the smaller ($D\lesssim200$~m) NEAs: This size regime has been poorly sampled; it is susceptible to radiation pressure effects, which could then be quantified; allows probing below the ``spin barrier'' at $\sim$2.2~h; can show if at these sizes, asteroids still have satellites; does the fraction of tumblers increase with size?

As most new discoveries will be smaller NEAs, they will likely be fast-spinning and fade fast, so cannot be observed for many nights. For light curve requirements for these objects we adopt as a benchmark the same photometric precision of 3\% and set the cadence at 30~s to potentially reach periods under 10~minutes (or less if several periods are observed). We conclude that even a 4~m-class telescope could accomplish this only for the brighter new discoveries on the same night.

Some light curve work has already been done on a very large telescope. \cite{Kwiatkowski2010a} reported on a pilot program to use the 10~m SALT telescope to measure 14 NEAs with $R\sim20$, $H>21.5$ with 5--60~s exposures (various instrumental issues limited them to $\sim$0.1~mag precision). They measured periods ranging from 77~s to 44~min. A dedicated campaign that observed all of the newly-discovered $H>22$ NEAs for 3~months could double the sample of periods for this population.  

Finally, \cite{Warner2011b} cite the potential usefulness (and outreach synergy) of smaller (0.5--1~m) telescopes in obtaining light curve observations in the range $V=17\text{--}19$. Exposure times to reach the required S/N would force a multi-rotation Fourier analysis approach for most of this range on small telescopes, which is better-suited to the large ($H\leq22$) NEAs.  Figure~\ref{dmaglines} indicates current rates of $\sim$1 new discovery per night overall within this magnitude range.  But for any newly-discovered $H>22$ NEAs (and of the known population, about 1/4 were discovered with $V<19$), their (likely) fast rotations would demand faster cadences and a 2~m-class telescope.  Meanwhile, for reapparitions of currently known NEAs, $\sim$10\% (or $\sim$1,000 NEAs) will achieve $V=18$ within a decade (\S\ref{subsequent}).

\subsection{Requirements for Spectra}
\label{spectroscopy}

Optical reflectance spectra with modest resolution ($R=100$) of Main Belt asteroids were the basis of the \cite{BusBinzel2002} asteroid taxonomy scheme. They provide a breakdown of types into Carbonaceous (C), Stony (S), and Metallic and Unknown (X), and into several sub-types. The currently standard optical-infrared Bus-DeMeo taxonomy \citep{DeMeo2009} is superior to the Bus-Binzel taxonomy as it adds spectral diagnostics in the near-IR (1.0--2.5~microns) to the optical.  All but three of the 26 Bus-Binzel taxonomy classes are preserved, with just one new class (Sv) defined in the Bus-DeMeo scheme. For some purposes, the visible-only regime preserves useful discriminants among types  \citep[e.g., see Figures 8, 10 and 13 in ][]{BusBinzel2002}. From the NEO perspective, many key classes are defined only by their visible-wavelength features; e.g., the selection of potential metallic M-type objects from the two other sub-classes of the X-class, is improved by rejecting the $\sim$30\% of E-types with a strong 0.49~mm optical feature. The 0.7~mm band, though uncommon in NEOs, is the only alternative to the 3.1~mm feature to diagnose hydrated minerals.

Optical to infrared spectra of NEAs as faint as $V=17.5$ can be obtained with an exposure of 30--60~minutes using the spectrograph SpeX on the 3~m Infrared Telescope Facility (IRTF) on Mauna Kea, Hawai'i\footnote{\url{http://irtfweb.ifa.hawaii.edu/~spex/}} (R.P. Binzel, priv. comm.).  Figure~\ref{dmaglines} indicates that this would apply to $\sim$100 ($\sim$10\% and decreasing) newly-discovered NEAs per year. Optical to infrared spectra of NEAs as faint as $V=20$ \citep{Moskovitz2012} can be taken in a 1~hour exposure with the FIRE spectrograph on the 6.5~m Magellan telescopes at Las Campanas Observatory in Chile\footnote{\url{http://obs.carnegiescience.edu/Magellan/}} \citep{Simcoe2010}. However, for projected new discoveries, $V=20$ ($R=19.5$) would reach only the brightest $\sim$1/3 of objects (see \S\ref{fiducial_magnitudes}), and then only on the discovery night.  Most $H\leq22$ reapparitions would be reachable, but only if the astrometry were of sufficient accuracy to be able to place the NEA in the spectrograph slit.

An alternative approach is to forego infrared spectroscopy. Optical spectroscopy is much less demanding than near-infrared spectroscopy as the atmospheric background is orders-of-magnitude fainter. A 2~m-class  telescope with an efficient spectrograph can obtain a modest resolution ($R=100$) spectrum with S/N $\sim$15 for a $V=20.5$ ($R=20$) NEA in under 1~h \citep[e.g., the LINNEAUS project; see][]{Elvis2014c}; this represents the mean new discovery on the discovery night, and most $H\leq22$ reapparitions.  New discoveries would require a 4~m telescope after a week. 

This argues strongly for a dedicated 2~m telescope equipped with a high-efficiency visible wavelength $R=100$ spectrometer. At a good site, a telescope can effectively observe faint NEAs for $\sim$210~nights of the year, after weather (15\%), full Moon (25\%) and equipment maintenance (5\%) are excluded\footnote{\url{http://www.gemini.edu/sciops/statistics/}}. Assuming an average $\sim$10~h of observation per night, a dedicated NEA telescope/spectrograph combination could obtain $\sim$2,000~spectra/year. Such a setup could keep pace with new NEA discoveries with same/next night spectroscopic followup. It would double the sample of NEA spectra within the first 6~months and would extend the coverage to small objects.

\subsection{Requirements for Colors}
\label{colors}

In the absence of spectra, broad-band colors can in some cases provide some level of discrimination among asteroid types, for example as shown for the griz SDSS filters by \cite{DeMeo2013} and \cite{Ivezic2001}. From their figures, we judge that photometry at the level of a few percent would be sufficient (higher precision would not necessarily be helpful as the scatter in colors within a population is typically on the order of 0.1~mag).  One significant complication is that the NEA's light in all bands will be modulated by its rotation.  For big NEAs ($H\leq22$), periods can be many hours, but for the smaller NEAs periods are typically $<$2.2~h (see \S\ref{light_curves}). For very fast rotators, long exposures would give an average over the light curve and thus could be used to get (average) colors.  But for exposures on, or separated by, timescales significant compared to the (generally unknown) period, there will be light-curve-dependent corrections depending on the (generally unknown) amplitude, which is often nearly half a magnitude (see \S\ref{light_curves}). In any case, exposures can be tracked on the asteroid.

Interpolating between the $\text{S/N}=10$ and 100 columns in Table~\ref{exp_time}, and using the $R$-band exposure time as a reasonable estimate for the average exposure time over $g$, $i$, and $z$, we conclude that 2~m telescopes could achieve the required precision in short (minutes per filter) exposures for most $H\leq22$ reapparitions and the mean new discoveries.  Longer exposures (hours to get multiple filters) would be needed for each new mean NEA a week after discovery, which would only be practical for fast rotators but which is not likely worth the time given the crude diagnostic they would permit; low-resolution spectroscopy as described above would be a better choice.  A 4~m telescope would be somewhat more practical for multicolor photometry of fast-rotating new discoveries out to a week, where 4 filters could be done on an hour timescale.

\subsection{Requirements for Phase Curves}
\label{phase_curves}

By measuring the brightness of an asteroid over a range of phase angles ($\alpha$) it is possible to derive an estimate of its absolute magnitude. Moreover, it has been shown that different taxonomic classes show distinct linear relationships between phase angle and magnitude (for $\alpha\gtrsim3$\degree), and that the slope of this line (the phase coefficient, $\beta$) is related to the albedo of the asteroid \citep{Belskaya2000}. These relationships appear to hold true for smaller asteroids too \citep{Hergenrother2013}. Thus, by measuring a phase curve we obtain an estimate of the asteroid's albedo and broad taxonomic class, as well as its absolute magnitude. If the absolute magnitude is already known, the phase curve can be used to estimate the slope parameter $G$.

10--20\% photometry would be good enough for constraining the phase curve if measured over a large enough range of phase angles, and hence over long enough observation arcs \citetext{C. Hergenrother, priv.\ comm.}. The photometric S/N constraint is more stringent than the minimal one for astrometry, and thus astrometric measurements could be made ``for free'' from these observations. Or, put another way, useful phase curve photometry could come out of the data taken for astrometry if the field were large enough to contain some of the growing number of faint photometric standard stars.  Given that there are already sufficient stars in a 30 arcmin field of view to permit photometry at the few percent level \citep{PicklesDepagne2010} and that the AAVSO Photometric All-Sky Survey\footnote{\url{http://www.aavso.org/apass/}} is still in progress, a 20~arcmin field like that suggested for astrometry will prove sufficient. The photometric standard surveys have a faint limit of $V\sim16\text{--}17$.  While exposures targeting the magnitudes expected for the faintest NEA discoveries might have these faint standards saturated, interweaving quick exposures would permit photometric calibration.  

From Table~\ref{exp_time} we conclude that a 2~m telescope could be used for phase curve measurements for most $H\leq22$ reapparitions as well as most new discoveries out to a couple of weeks. A 4~m could extend coverage to a month for the mean new discovery at an hour per exposure (but keep in mind the complications from light curve modulation; see \S\ref{colors}). This might not be justifiable based on the phase curve knowledge alone but it would make sense to plan the long-arc astrometric followup observations to have longer exposure times (either in single exposures or by coadding) than strictly necessary for the astrometry itself, to permit the phase curve measurements at large phase angles.

%%%%%%%%%%%%%%%%%%%% CONCLUSIONS
\section{Conclusions} 
\label{conclusions}

We have used properties of the over 10,000 known NEAs and their discovery circumstances to show that rapid follow-up after discovery is essential for closing the wide, and increasing, gap between discovery and characterization on a decade timescale, particularly as an increasing number of new discoveries will be of smaller ($H>22$) NEAs that are projected to be 100$\times$ fainter on subsequent apparitions.  

Spectral follow-up that keeps pace with discoveries could be accomplished with a dedicated 2~m telescope observing NEAs within days of discovery. Long-arc (months) astrometry would require a 4+~m telescope, and the same observations could be used to measure phase curves and colors that would help to constrain albedos. Light curves for the bigger (and likely more slowly-rotating) new NEAs could be collected shortly after discovery with a 2~m telescope, but the cadence required to measure fast rotations likely for smaller new NEAs would necessitate rapid follow-up with a 4~m telescope.

Already-known large ($H\leq22$) NEAs making return apparitions are predicted to reach brightnesses similar to their discovery magnitudes (generally brighter than the discoveries projected for new surveys, which will thus likely measure astrometry for them by default).  For this subset of objects, targeted follow-up spectroscopy and photometry can be done with 2~m telescopes. For the brightest of these, 1~m-class amateur telescopes can contribute light curves and astrometry.

If means are insufficient to fund the necessary telescopes, which should be prioritized? Spectral classification yields the highest scientific return as it provides composition, which also yields an albedo estimate, and thus a more stringent constraint on an object's size. Light curves may provide structural information or reveal a binary system.

Lastly, astrometric measurements are of value if one desires to perform targeted observations at a subsequent apparition after discovery and thus requires a precise orbit. It should be noted that orbits obtained within the first few days or weeks of an NEA's discovery are generally good enough to securely establish the shape and location of the orbit in space \citep{Lee2015}, while further observations, especially at future aparitions, will do most to improve the accuracy of locating the asteroid along its orbit, and thus on the sky. The level of accuracy achieved for the initial orbit is sufficient for studies of the shapes and distribution of asteroid orbits and their evolution over time; it is also sufficient to determine if it intersects with, or is close enough to, Earth's orbit that an asteroid with that orbit could be at risk of impact with our planet.

Because the majority of new NEAs are not going to pose any danger of short- or mid-term impact, it may be argued that a more precise orbit at time of discovery is unnecessary because an NEA will be serendipitously reobserved by the surveys at some future apparition(s) and linked back by the Minor Planet Center to its discovery apparition. In this case, improving the accuracy of an orbit becomes a game of wait-n'-see, which argues strongly for characterization to take place during the discovery apparition as there is no guarantee a new NEA will be found at a later apparition with a targeted observation. This strategy, however, is unacceptable for mission planning, where the location of a target asteroid would need to be known with high accuracy years ahead of launch. Commercial ventures planning to send probes to many NEAs of varying sizes \citep{Elvis2014b} would benefit greatly from more accurate orbits being determined during the discovery apparition if they intend to be able to act quickly on newly-discovered profitable asteroids.

We have shown that a small group of dedicated telescopes (one 4~m and two 2~m, with help from smaller amateur contributions) could characterize or constrain the spectral class, albedo and spin period of most new NEA discoveries while also working through the backlog of known NEAs. There are compelling scientific reasons to do, which also have implications for planetary defence: Understanding the composition and structural make up of the smaller ($H\leq22$) NEAs will provide knowledge of the size group of asteroids most likely to impact Earth in the near future, while testing current theories of how the NEA population is fed by the Main Belt. 

%%%%%%%%%%%%%%%%%%%% ACKNOWLEDGMENTS
\section*{Acknowledgments}

We thank both T. Spahr and G. Williams (MPC) for answering numerous questions, and the anonymous referees for keeping us honest and improving this paper through their inquisitiveness. C. Beeson thanks M. Trichas and J. Connolly for their guidance with Python coding; D. Polishook for an insight into NEA spectroscopy; and as F.E. DeMeo and R.P. Binzel for crucial information with regards to spectroscopy. J.L. Galache was funded by NASA Grant NNX12AE89G. This research has made use of data and/or services provided by the International Astronomical Union's Minor Planet Center.

%%%%%%%%%%%%%%%%%%%% REFERENCES


\bibliographystyle{elsarticle-harv}
\bibliography{galache-asteroid-bibliography}
%% not want to use model2-names.bst.

\end{document}